\documentclass[12pt]{article}
\usepackage{amssymb}

\newcommand{\be}{\begin{equation}}
\newcommand{\ee}{\end{equation}}
\def\bea{\begin{eqnarray}}
\def\eea{\end{eqnarray}}

\newcommand{\bn}{\begin{eqnarray}}
\newcommand{\en}{\end{eqnarray}}

\newcommand{\p}{\partial}

\newcommand{\nn}{\nonumber}

\newcommand{\no}{\noindent}

\def\bea{\begin{eqnarray}}
\def\eea{\end{eqnarray}}

\newcommand{\beq}{\begin{eqnarray}}
\newcommand{\eeq}{\end{eqnarray}}

\begin{document}

\title{\textbf{Hamiltonian analysis and positivity of a new massive spin-2 model}}
\author{{ Alessandro L. R. dos Santos $^{1}$\footnote{alessandroluiz1905@gmail.com} , Denis Dalmazi $^{2}$\footnote{denis.dalmazi@unesp.br} ,
Wayne de Paula $^{1}$\footnote{lspwayne@gmail.com}}\\
\textit{{1- Instituto Tecnol\'ogico de Aeron\'autica, DCTA} }\\
\textit{{CEP 12228-900, S\~ao Jos\'e dos Campos - SP - Brazil} }\\
\textit{{2- UNESP - Campus de Guaratinguet\'a - DFI} }\\
\textit{{CEP 12516-410, Guaratinguet\'a - SP - Brazil.} }\\}
\date{\today}
\maketitle

\begin{abstract}
Recently a new model has been proposed to describe free massive spin-2 particles in $D$ dimensions
in terms of a non symmetric rank-2 tensor $e_{\mu\nu}$ and a mixed symmetry tensor $B^{\mu[\alpha\beta]}$.
The model is invariant under linearized diffeomorphisms without Stueckelberg fields.
It resembles a spin-2 version of the topologically massive spin-1 BF model (Cremmer-Scherk model).
Here we apply the Dirac-Bergmann procedure in order to identify all Hamiltonian constraints
and perform a  complete counting of degrees of freedom.
In $D=3+1$ we find $5$ degrees of freedom corresponding to helicities $\pm{2},\pm{1},0$ as expected.
The positivity of the reduced Hamiltonian is proved by using spin projection operators.
We have also proposed a parent action that establishes the duality between the Fierz-Pauli and
the new model. The equivalence between gauge invariant correlation functions of both theories is demonstrated.

\end{abstract}

\newpage

\section{Introduction}


The experimental data based on supernova measurements \cite{Riess,Perlmutter} have indicated an accelerated expansion of the Universe at large distances. The Einstein's general relativity theory explain this phenomenon by introducing a cosmological constant $\Lambda$ that works like a constant energy density called Dark Energy. Attempts to modify  gravity at large distances by means of a massive graviton provide an interesting alternative explanation for the observed acceleration of the Universe without Dark Energy, see \cite{deRham-1, Hinter-1}. An acceleration driven by a small graviton mass might be more natural than a small cosmological constant, see comments in \cite{hinterc}.   Another important effect of a massive graviton is the presence of extra states of polarization modes for gravitational waves \cite{dePaula}.
In \cite{Damico} an upper limit for the graviton mass was estimated to be an order of magnitude below the present-day value of the Hubble parameter in the low density regime of the Universe. On the other hand, the recent detections of gravitational waves \cite{Ligo} have fixed an upper limit for the graviton mass in $10^{-23}$ eV; therefore the existence of massive graviton is not ruled out. From a theoretical point of view, massive gravitons have been investigated in the past, built up on the top of the free massive spin-2 Fierz-Pauli (FP) model \cite{Fierz}. At that time, two problems were raised: the discontinuity of the massless limit (vDVZ discontinuity)\cite{vDV,Zak} and the appearance of a nonphysical degree of freedom (Boulware-Deser ghost) in the nonlinear theory \cite{bd}. The first problem was overcome in \cite{vainsh} by taking into account nonlinear terms (Vainshtein mechanism), while the second problem was solved in \cite{deRham-2,rosen} by a judicious choice of a non-linear potential for the metric fluctuation $h_{\mu\nu}$.
The model of \cite{deRham-2} is known as the dRGT model.

The dRGT model contains a fixed Minkowski metric besides the dynamical metric. After expanding the dynamical metric about a fixed background the dRGT model at the linear level reduces to the FP model. The reparametrization symmetry is only achieved by adding Stueckelberg fields. The fixed Minkowski metric of the dRGT model was generalized to an arbitrary fixed metric in \cite{hrnl}. A later improvement, based on the non linear potential found in \cite{deRham-2}, is the  bimetric model of \cite{hr} where the fixed metric is promoted to a dynamical one and the model is now manifestly diffeomorphisms invariant and accommodate viable FLRW cosmological solutions, see \cite{hogas1,hogas2} for a recent comparison with observational tests and further comments. The price we pay is that the model now contains two gravitons,  a massive  and a massless one. Since the bimetric model is also built upon the Fierz-Pauli theory, it is natural to ask for possible dual models to the free massive FP theory  which could be used as an alternative starting point for building up diffeomorphism invariant massive gravity models without introducing extra propagating modes.

Recently, in \cite{dual} we have proposed a new massive spin-2 model dual to the FP theory. It is  similar to the $U(1)$ invariant Cremmer-Scherk model \cite{cremmer}, also called topologically massive BF model in $D=3+1$,  which describes massive spin-1 particles in a gauge invariant way and contains a non Abelian counterpart \cite{ft}. The massive model presented in \cite{dual} is given in terms of a non-symmetric rank-2 tensor $e_{\mu\nu}$ coupled to a mixed symmetry tensor $B^{\mu[\alpha\beta]}$ similar to a vielbein and a spin connection respectively. The model is invariant under linearized diffeomorphism. However, since the algebraic manipulations performed in \cite{dual} involve field redefinitions with time derivatives, it is not clear whether the canonical structure of the phase space  is completely faithful to massive spin-2 particles. Moreover, for possible nonlinear extensions and ghost free proofs the knowledge of the basic canonical structure is essential \cite{rosen,kluson}.  Here we use the Hamiltonian approach in the proof of absence of ghosts in this dual model. Applying the Dirac procedure \cite{Dirac,Henneaux} we identify all Hamiltonian constraints and demonstrate that the counting of degrees of freedom is consistent with only massive spin-2 particles in the spectrum. We use spin projection operators to prove that the reduced Hamiltonian is definite positive and that all expected helicities ($\pm 2,\pm 1, 0$), in $D=3+1$, are present. In addition, we suggest a simpler parent action than the one in \cite{dual}, that allows us interpolating between the FP model and the dual massive spin-2 model. We also point out that the model of \cite{dual} in $D=2+1$ is a kind of ``zwei-dreibein'' model \cite{zwei}.

In section 2 we carry out the Hamiltonian analysis of the spin-1 massive BF model in $D$ dimensions  establishing our setup for future comparison with the spin-2 case. The Hamiltonian analysis  of the massive BF model  in $D=3+1$ dimensions appears in \cite{Lahiri}.
In the section 3 we introduce the parent action that allows interpolating between the FP and the dual massive spin-2 model.
Then we apply the Hamiltonian analysis in this dual model and check its positivity. In section 4 we present our conclusions and perspectives.
In the appendix we present the basic formulae for the spin projection and transition operators.

\section{Massive spin-1 BF model}

Massive spin-1 particle are commonly described by the Proca model, whose mass term breaks symmetry under $U(1)$ transformations.
An invariant gauge formulation for massive spin-1 is known as the BF model (or Cremmer-Scherk model) \cite{cremmer} which has a nonabelian extension \cite{FT}. This model depends on the vector field $A_{\mu}$ and an auxiliary antisymmetric tensor field $B^{\mu\nu}$. In $D$ dimensions the duality between the Proca and BF models can be easily verified through the parent action

\bea
S[J] = \int{d}^{D}x\Bigg\{-\frac{1}{4}F^{\,2}_{\mu\nu}(A)- \frac{m^{2}}{2}V^{\mu}V_{\mu}+\frac{m}{2}B^{\mu\nu}F_{\mu\nu}(V - A)+V_{\mu}J^{\mu}\Bigg\}
\label{parent-s1}\;, \eea
where $V^{\mu}$ is an auxiliary vector field and $F_{\mu\nu}(A)$ is the electromagnetic field strength defined by $F_{\mu\nu}(A)=\partial_{\mu}A_{\nu}-\partial_{\nu}A_{\mu}$. We have also added a symmetric external source $J^{\mu}$.
This action is invariant (up to a total derivative) under two independent gauge transformations.

\bea
\delta_{\varphi}A_{\mu} = \partial_{\mu}\varphi \qquad,\qquad \delta_{\Omega}B_{\mu\nu} = \partial^{\alpha}\Omega_{[\alpha\mu\nu]}\;,
\label{BF-symm}
\eea
\no where $\Omega_{[\alpha\mu\nu]}$ is a fully antisymmetric tensor \footnote{In this work we use $\eta_{\mu\nu}=(-,+,+,\ldots,+)$. In addition $e_{(\mu\nu)}=(e_{\mu\nu}+e_{\nu\mu})/2$ , $e_{[\mu\nu]}=(e_{\mu\nu}-e_{\nu\mu})/2$ , and so on.}.

Let us define the generating function $Z[J]$ which will allow us to derive a dual map between correlation functions in the dual theories

\bea
Z[J] = \int\;\mathcal{D}A_{\mu}\mathcal{D}B_{\alpha\beta}\mathcal{D}V_{\nu}\;e^{\;i\,S[J]} \,.
\label{Z-1}\eea
The Gaussian integration over the field $B^{\mu\nu}$ results in the functional constraint $F_{\mu\nu}(V-A)=0$ whose general solution is $V_{\mu}=A_{\mu}+\frac{\partial_{\mu}\psi}{m}$, where $\psi$ is an arbitrary scalar field. Back in action we obtain the Stueckelberg-like Proca model theory with source

\bea
S_{Proca}[J] = \int{d}^{D}x\Bigg\{-\frac{1}{2}F^{\,2}_{\mu\nu}(A)-\frac{m^{2}}{2}\Big(A_{\mu}+\frac{\partial_{\mu}\psi}{m}\Big)^{2}+\Big(A_{\mu}+\frac{\partial_{\mu}\psi}{m}\Big)J^{\mu}\Bigg\}\;,
\label{proca} \eea
which is invariant under the transformations $\delta_{\varphi}A_{\mu}=\partial_{\mu}\varphi$ and $\delta_{\varphi}\psi=-m\varphi$, where $\psi$ act a Stueckelberg field.

On the other hand, integrating over the auxiliary field $V_{\mu}$ and performing the shift $V_{\mu}\,\rightarrow\,V_{\mu}-\frac{\partial^{\nu}B_{\mu\nu}}{m}+J_{\mu}$ we obtain the BF model with source

\bea
S_{BF}[J] &=& \int{d}^{D}x\Bigg\{-\frac{1}{4}F^{\,2}_{\mu\nu}(A)+\frac{1}{2}\partial_{\alpha}B^{\alpha\mu}\partial^{\beta}B_{\beta\mu}-\frac{m}{2}B^{\mu\nu}F_{\mu\nu}(A)\nn\\
&&\qquad\qquad-\,\frac{1}{m}\partial^{\alpha}B_{\alpha\mu}J^{\mu}+\mathcal{O}(J^{\,2})\Bigg\}\;.
\label{BF} \eea
In $D=3+1$ dimensions we can redefine $B^{\mu\nu}=\epsilon^{\mu\nu\alpha\beta}\widehat{B}_{\alpha\beta}$ and the model is known as topologically massive BF model. The action (\ref{BF}) is invariant under (\ref{BF-symm}) without Stueckelberg-like fields.

The equations of motion from (\ref{BF}) at vanishing source ($J^{\mu}=0$) can be written as

\bea
 F^{\mu\nu} = \partial^{\mu}\mathbb{B}^{\nu}-\partial^{\nu}\mathbb{B}^{\mu}  \label{BF-eq-2}\;,
\eea
\bea
\partial_{\nu}F^{\nu\mu} = m^{2}\mathbb{B}^{\mu} \label{eq-1} \label{BF-eq-1}\;,
\eea
where $\mathbb{B}^{\mu} = -\,\partial_{\nu}B^{\nu\mu}/m\,$ is a gauge invariant quantity. Substituting Eq.(\ref{BF-eq-2}) in Eq.(\ref{BF-eq-1}) we obtain the Klein-Gordon equation

\bea
(\square - m^{2})\mathbb{B}^{\nu} = 0 \;. \label{KG-s1}
\eea
Note that the transversality condition is an identity: $\partial_{\mu}\mathbb{B}^{\mu}=-\partial_{\mu}\partial_{\nu}B^{\mu\nu}/m=0$, while in the Proca model the transversality condition $\partial_{\mu}A^{\mu}=0$ follow from the equation of motion. This is peculiar to dual theories.  The transversality condition and Klein-Gordon equation are all we need to describe free massive spin-1 particles.

Let us obtain the quantum equivalence between correlation functions by deriving $Z[J]$ with respect to $J^{\mu}$ for (\ref{proca}) and (\ref{BF}) obtaining

\bea
\Big\langle\widetilde{A}_{\mu_{1}}(x_{1})\ldots\widetilde{A}_{\mu_{N}}(x_{N})\Big\rangle_{Proca} =\Big\langle\mathbb{B}_{\mu_{1}}(x_{1})\ldots\mathbb{B}_{\mu_{N}}(x_{N})\Big\rangle_{BF} + \mbox{contact term}\;, \label{ct}
\eea
where $\widetilde{A}_{\mu}=A_{\mu}+\frac{\partial_{\mu}\varphi}{m}$.
So we have the following gauge invariant map between $S_{Proca}$ and $S_{BF}$

\bea
\widetilde{A}_{\mu}\;\leftrightarrow\;\mathbb{B}_{\mu}\;. \label{map1}
\eea

\no Notice that the contact terms appear due to the quadratic terms in the sources in (\ref{BF}). In (\ref{ct}) and also in (29) the presence of contact terms shows that dual maps like  (\ref{map1}) and (30) can not be implemented directly in the action where we have products of fields at the same space-time point, there would be no equivalence of the actions, see discussion in \cite{renato2}.

\subsection{Hamiltonian analysis}

We start the Hamiltonian analysis in $D$ dimensions by separating the time and space components in the BF Lagrangian density. Up to total integrations we obtain

\bea
\mathcal{L}_{BF} &=&
\frac{1}{2}\dot{A}_{i}\dot{A}_{i}+\frac{1}{2}\dot{B}_{0i}\dot{B}_{0i} - \dot{A}_{i}\partial_{i}A_{0}-\dot{B}_{0i}\partial_{j}B_{ji}-m\dot{B}_{0i}A_{i}+\frac{1}{2}\partial_{i}A_{0}\partial_{i}A_{0}\nn\\
&&-\,\frac{1}{2}F^{\,2}_{ij}(A)-\frac{1}{2}\partial_{i}B_{0i}\partial_{j}B_{0j}+\frac{1}{2}\partial_{i}B_{ik}\partial_{j}B_{jk}-m\,B_{0i}\partial_{i}A_{0}-\frac{m}{2}B_{ij}F_{ij}(A)\;.
\eea
From the canonical momenta $\pi^{\mu} = \delta\mathcal{L}_{BF}/\delta\dot{A}_{\mu} $ and $\Pi^{\mu\nu} = \delta\mathcal{L}_{BF}/\delta\dot{B}_{\mu\nu}$ we obtain the following primary constraints

\bea
\varphi = \pi^{0} \approx0 \qquad \mbox{and} \qquad \Phi^{ij} = \Pi^{ij} \approx0 \;.
\label{primary-s1}
\eea
Using the Lagrange multipliers $\lambda$ and $\lambda_{ij}$ we add the primary constraints (\ref{primary-s1}) to canonical Hamiltonian $H=\int{d}^{D-1}x\,\mathcal{H}$ in order to obtain the primary Hamiltonian

\bea
H_{P} = \int{d}^{D-1}x\Big[\mathcal{H}+\lambda\varphi+\lambda_{ij}\Phi^{ij}\Big] \;,
\label{HP-s1}
\eea
where $\mathcal{H}$ is the canonical Hamiltonian density,

\bea
\mathcal{H}&=&\pi^{\mu}\dot{A}_{\mu}+\Pi^{\mu\nu}\dot{B}_{\mu\nu}-\mathcal{L}_{BF} \nn\\
&=&\frac{1}{2}\pi^{i}\pi^{i}+\frac{1}{2}\Pi^{0i}\Pi^{0i}+\pi^{i}\partial_{i}A_{0}+m\Pi^{0i}A_{i}+\Pi^{0i}\partial_{j}B_{ji} \nn\\
&&+\frac{m^{2}}{2}A_{i}A_{i}+\frac{1}{4}F^{\,2}_{ij}(A)+\frac{1}{2}\partial_{i}B_{0i}\partial_{j}B_{0j}+m\,B_{0i}\partial_{i}A_{0}\;.
\label{density-s1} \eea

Applying the consistency condition to the primary constraints we find the secondary constraints $\chi$ and $\chi^{ij}$

\bea
\dot\varphi &=& \{\varphi\,,\,H_{P}\}\approx0 \qquad \rightarrow \qquad
\chi = \partial_{i}\pi^{i}+m\,\partial_{i}B_{0i}\;\approx\;0 \;, \label{secondary-s1}\\
\dot\Phi^{ij} &=& \{\Phi^{ij}\,,\,H_{P}\}\approx0 \qquad \rightarrow \qquad
\chi^{ij} = \partial_{i}\Pi^{0j}-\partial_{j}\Pi^{0i}\;\approx\;0 \;. \label{secondary-s11}
\eea

In turn the secondary constraints (\ref{secondary-s1}-\ref{secondary-s11}) identically satisfy the consistency conditions and there are no tertiary constraints. All constraints are of first class since the Poisson brackets between them are weakly  zero. It is postulated, in general, that all first class constraints generate gauge transformations \footnote{Dirac's conjecture states that all first-class constraints generate symmetries, but there are counterexamples \cite{Henneaux}.}. When the Poisson bracket between two constraints is non-zero, both constraints are of the second class. The Lagrange multipliers remain undetermined since the number of undetermined multipliers is equal to the number of first class primary constraints.

The BF model written in terms of  $A_{\mu}$, $B^{\mu\nu}$ and their momenta has $N=D(D+1)$ independent components in the phase space.
On the other hand the constraints $(\pi^0\,,\,\chi\,,\,\Pi^{ij}\,,\,\chi^{ij})$ correspond respectively to $(1\,,\,1\,,\,(D-1)(D-2)/2\,,\,D-2)$ constraints\footnote{Notice that in general an antisymmetric tensor like  $\chi^{ij}$ has $(D-1)(D-2)/2$ components, however our constraints $\chi^{ij}=0$ lead in general to $\Pi^{0j}=\p^j\phi$, thus reducing $D-1$  to 1 degree of freedom which means that we have $D-2$ independent constraints. In particular, in $D=4$ we have the identity $\epsilon_{kij}\p^k\chi^{ij}=0$ which reduces 3  to 2 constraints.}, thus adding up to  $(D^{2}-D+2)/2$ first class constraints (FCC). The number of degrees of freedom ($dof$) in the phase space is given by

\bea
dof = N - 2\times{FCC} = 2(D-1) \;.
\eea
In agreement with  the three helicities ($\pm 1$, $0$) in $D=3+1$ and
the two helicities ($\pm 1$) in $D=2+1$.

In order to check the positivity of the reduced Hamiltonian, we use the constraints (\ref{primary-s1}) and
(\ref{secondary-s1}-\ref{secondary-s11}) as strong equalities in the primary Hamiltonian (\ref{HP-s1}). This results in the reduced definite positive Hamiltonian

\bea
H^{(r)}_{P} &=& \int{d}^{D-1}x\Bigg[\frac{1}{2m^{2}}(\partial_{i}\pi^{i})^{2}+\frac{1}{2}\pi^{i}\pi^{i}+\frac{1}{2}F^{\,2}_{ij}(A)+\frac{m^{2}}{2}\Big(A_{i}+\frac{\Pi^{0i}}{m}\Big)^{\,2}\Bigg]\;. \label{hpr}
\eea
Note that (\ref{hpr}) coincides with the Proca reduced hamiltonian after a canonical transformation $\widehat{A}_{i}=A_{i}+\frac{\Pi^{0i}}{m}$ and $\widehat{B}_{0i}=B_{0i}+\frac{\pi^{i}}{m}$.

\section{Dual massive spin-2 model}

In \cite{dual} a new massive spin-2 model was derived from a massive parent action which in turn was obtained via Kaluza-Klein dimensional reduction of a dual first order version to Vasiliev's action for massless spin-2 particles (linearized first order Einstein-Hilbert) \cite{vasiliev}. The new model is invariant under two independent gauge transformations and it can be considered the generalization of the BF action to spin-2. Here we present an alternative and simplified way to demonstrate the duality between the new massive spin-2 and the FP models. Consider the following parent action with source term (compare with (\ref{parent-s1}))

\bea
S[J] = \int{d}^{D}x\Bigg\{\mathcal{L}_{LEH}(e)-\frac{m^{2}}{2}(W^{\mu\nu}W_{\nu\mu}-W^{2})
+\frac{m}{2}B^{\mu[\alpha\beta]}F_{[\alpha\beta]\mu}(W-e)+W_{\mu\nu}J^{\nu\mu}\Bigg\}\;,
\label{parent-s2}\eea
where $W_{\mu\nu}\ne W_{\nu\mu}$ are auxiliary fields, with $W=\eta^{\mu\nu}W_{\mu\nu}$. The field strenght $F_{[\alpha\beta]\mu}$ is defined by
$F_{[\alpha\beta]\mu}(e) = \partial_{\alpha}e_{\beta\mu}-\partial_{\beta}e_{\alpha\mu}$ and $\mathcal{L}_{LEH}(e)$ is the linearized Einstein-Hilbert theory

\bea
\mathcal{L}_{LEH}(e)=-\frac{1}{8}F^{[\alpha\beta]\mu}F_{[\alpha\beta]\mu}-\frac{1}{4}F^{[\alpha\beta]\mu}F_{[\alpha\mu]\beta}+\frac{1}{2}F^{\alpha}F_{\alpha}\;,
\eea
where $F^{\alpha}=\eta_{\mu\beta}F^{[\alpha\beta]\mu}$. The parent action (\ref{parent-s2}) is invariant under the independent transformations

\bea
\delta_{\xi}e_{\mu\nu}=\partial_{\mu}\xi_{\nu} \qquad,\qquad \delta_{\Omega}B^{\mu[\alpha\beta]} = \partial_{\nu}\Omega^{\mu[\nu\alpha\beta]}\;.
\label{symm-s2} \eea

We can define the generating function

\bea
Z[J] = \int\;\mathcal{D}e_{\mu\nu}\mathcal{D}W_{\alpha\beta}\mathcal{D}B_{\rho[\lambda\sigma]}\;e^{\,i\,S[J]}\;.
\label{Z-2}\eea

The integral over $B^{\mu[\alpha\beta]}$ results in the functional constraint $F_{[\alpha\beta]\mu}(W-e)=0$ whose general solution is $W_{\mu\nu}=e_{\mu\nu}+\frac{\partial_{\mu}A_{\nu}}{m}$, where $A_{\mu}$ is an arbitrary vector field. Back in (\ref{parent-s2}) we obtain the Stueckelberg-like Fierz-Pauli model with source

\bea
S_{FP} &=& \int{d}^{D}x\Bigg\{\mathcal{L}_{LEH} - \frac{m^{2}}{2}\Big(e_{\mu\nu}+\frac{\partial_{\mu}A_{\nu}}{m}\Big)\Big(e^{\nu\mu}+\frac{\partial^{\nu}A^{\mu}}{m}\Big)+\frac{m^{2}}{2}\Big(e+\frac{\partial^{\alpha}A_{\alpha}}{m}\Big)^{2} \nn\\
&&\qquad\qquad+\Big(e_{\mu\nu}+\frac{\partial_{\mu}A_{\nu}}{m}\Big)J^{\nu\mu}\Bigg\}\;.
\label{FP} \eea
It is invariant under the transformations $\delta_{\xi}e_{\mu\nu}=\partial_{\mu}\xi_{\nu}$ and $\delta_{\xi}A_{\mu}=-m\xi_{\mu}$.

On the other hand, integrating over $W_{\mu\nu}$ we obtain (after the shift in $W_{\mu\nu}$)  the dual massive spin-2 model proposed in \cite{dual} with sources\footnote{In \cite{ks} another model dual to the massive FP theory is suggested which contains besides a symmetric rank-2 tensor, an auxiliary B-field of rank-4. The model is of fourth order in derivatives.}

\bea
S_{Dual}[J]&=& \int{d}^{D}x\Bigg\{\mathcal{L}_{LEH}(e)+\frac{1}{2}\partial_{\alpha}B^{\mu[\alpha\nu]}\partial^{\beta}B_{\nu[\beta\mu]}-\frac{(\partial_{\alpha}B^{\alpha})^{2}}{2(D-1)}
-\frac{m}{2}B^{\mu[\alpha\beta]}F_{[\alpha\beta]\mu}(e)\nn\\
&&\qquad\qquad-\,\frac{1}{m}\Big[\partial^{\alpha}B_{\mu[\alpha\nu]}-\frac{\eta_{\mu\nu}\partial^{\alpha}B_{\alpha}}{(D-1)}\Big]J^{\nu\mu}+\mathcal{O}(J^{\,2})\Bigg\}\;, \label{BT}
\eea
where $B^{\alpha}=\eta_{\mu\beta}B^{\mu[\alpha\beta]}$. It is clearly invariant under the independent transformations (\ref{symm-s2}). The linearized Einstein-Hilbert theory $\mathcal{L}_{LEH}(e)$ describes by itself massless spin-2 particles while the kinetic term for the $B^{\mu[\alpha\beta]}$ field has no particle content as shown in \cite{bfmrt,renato2} and it is the same one appearing in the earlier massive spin-2 model of \cite{Curt}.

In terms of the invariant field strength $F^{\mu[\alpha\beta]}(e)$ and the invariant quantity

\bea
\mathbb{B}_{\mu\nu}=-\,\frac{\partial^{\alpha}B_{\mu[\alpha\nu]}}{m}+\frac{\eta_{\mu\nu}\partial^{\alpha}B_{\alpha}}{m(D-1)}\;,
\eea
we can write down the equations of motion from the action (\ref{BT}) at vanishing sources ($J^{\mu\nu}=0$) as

\bea F^{[\alpha\beta]\mu} = \partial^{\alpha}\mathbb{B}^{\beta\mu}-\partial^{\beta}\mathbb{B}^{\alpha\mu}\;,
\label{eom1}\eea
\bea \partial_{\alpha}F^{[\alpha\nu]\mu}+\partial_{\alpha}F^{[\alpha\mu]\nu}+\partial_{\alpha}F^{[\mu\nu]\alpha}-2\eta^{\mu\nu}\partial_{\alpha}F^{\alpha}+2\partial^{\mu}F^{\nu}
=2m^{2}[\mathbb{B}^{\mu\nu}-\eta^{\mu\nu}\mathbb{B}]\;,
\label{eom2}\eea
where $F^{\alpha}=\eta_{\mu\beta}F^{[\alpha\beta]\mu}$ and $\mathbb{B}=\eta_{\mu\nu}\mathbb{B}^{\mu\nu}$. We have shown in \cite{dual} that the equations of motion lead to the Fierz-Pauli \cite{Fierz} conditions and the Klein-Gordon equations

\bea
\mathbb{B} = \eta^{\mu\nu}\mathbb{B}_{\mu\nu} = 0 \quad,\quad \partial^{\mu}\mathbb{B}_{\mu\nu} = 0 \quad,\quad
\mathbb{B}_{[\mu\nu]} = 0 \quad,\quad(\square-m^{2})\mathbb{B}_{(\mu\nu)}=0 \;.
\label{kg} \eea
It is all we need to describe free massive spin-2 particles.

Deriving the generating function $Z[J]$ with respect to the source from (\ref{Z-2}) and considering (\ref{FP}) and (\ref{BT}) we obtain

\bea
\Big\langle\widetilde{e}_{\mu_{1}\nu_{1}}(x_{1})\ldots\widetilde{e}_{\mu_{N}\nu_{N}}(x_{N})\Big\rangle_{FP} =\Big\langle\mathbb{B}_{\mu_{1}\nu_{1}}(x_{1})\ldots\mathbb{B}_{\mu_{N}\nu_{N}}(x_{N})\Big\rangle_{Dual} + \mbox{contact term}\;,
\eea
where $\widetilde{e}_{\mu\nu}=e_{\mu\nu}+\frac{\partial_{\mu}A_{\nu}}{m}$. Therefore, we have the following dual map between gauge invariants

\bea
\widetilde{e}_{\mu\nu}\quad \leftrightarrow \quad \mathbb{B}_{\mu\nu}\;.
\eea

The equations (\ref{eom1}) give $F_{[\alpha\beta]\mu}$ in terms of $\mathbb{B}_{(\mu\nu)}$. As in the spin-1 case, since the particle content of the model can be encoded in $\mathbb{B}^{\mu\nu}$ which involves time derivatives of fundamental fields, it is not clear  whether the canonical structure from (\ref{BT}) is the correct one for massive spin-2 particles. So in the next section we check the constraints structure, the number of physical degrees of freedom and the Hamiltonian positivity. However, before we go on we mention the special case of $D=2+1$ where the particle content of the model can be easily established.

In $D=2+1$ we can rewrite (\ref{BT}) in terms of two arbitrary rank-2 tensors since, without loss of generality, we can always write $B^{\mu[\alpha\beta]}=\epsilon^{\alpha\beta\rho}f_{\rho}^{\;\;\mu}$. Back in (\ref{BT}), after using an $\epsilon$-identity\footnote{\bea \epsilon_{\mu\nu\lambda}\epsilon_{\alpha\beta\gamma} &=& -\eta_{\mu\alpha}(\eta_{\nu\beta}\eta_{\lambda\gamma}-\eta_{\nu\gamma}\eta_{\lambda\beta})-
\eta_{\mu\beta}(\eta_{\nu\gamma}\eta_{\lambda\alpha}-\eta_{\nu\alpha}\eta_{\lambda\gamma})- \eta_{\mu\gamma}(\eta_{\nu\alpha}\eta_{\lambda\beta}-\eta_{\nu\beta}\eta_{\lambda\alpha})\nn\eea.} we obtain

\bea
\mathcal{L}(e,B) = \mathcal{L}(e,f)= \mathcal{L}_{LEH}(e) + \mathcal{L}_{LEH}(f) +
m\,\epsilon^{\mu\nu\alpha}f_{\mu\beta}\,\partial_{\alpha}\,e_{\nu}^{\;\;\beta} \;.
\label{lef1} \eea
Notice that there is no propagating degree of freedom in the $D=2+1$ Einstein-Hilbert theory in agreement with the absence of particle content in the kinetic term for the $B^{\mu[\alpha\beta]}$ field in arbitrary dimensions. The model (\ref{lef1}), when written in first order, is a kind of ``zwei dreibein'' model, see \cite{zwei}. After a simple rotation we can decouple the fields and rewrite (\ref{lef1}) in terms of  two second order spin-2 self-dual models (parity singlets) of opposite helicities $+2$ and $-2$

\bea
\mathcal{L}(e,f)= \mathcal{L}_{SD2}^{(2)}(e^{+},m) + \mathcal{L}_{SD2}^{(2)}(e^{-},-m)\;,
\label{lef2}  \eea
where $e_{\mu\nu}^{\pm} = (e_{\mu\nu}\pm f_{\mu\nu})/\sqrt{2}$ and

\bea
\mathcal{L}_{SD2}^{(2)}(e,m) = \mathcal{L}_{LEH}(e) + \frac{m}{2}\,\epsilon^{\mu\nu\alpha}e_{\mu\beta}\partial_{\alpha}e_{\nu}^{\;\;\beta}\;,
\label{lsd2} \eea
is the spin-2 self-dual model of \cite{desermc}. It describes particles of helicity $2\,\vert m\vert/m$ without ghosts. Therefore, we do have two physical degrees of freedom ( helicities $\pm 2$) for $\mathcal{L}(e,B)$ in $D=2+1$, which is the same content of the massive FP theory in $D=2+1$.

\subsection{Hamiltonian constraints}

Splitting the Lagrangian density $\mathcal{L}(e,B)$ in time and space components we have (up to total derivatives)

\bea
\mathcal{L}(B,e) &=& \frac{1}{2}\dot{h}_{ij}\dot{h}_{ij}-\frac{1}{2}\dot{h}_{ii}\dot{h}_{jj}+\frac{1}{2}\dot{B}_{i[0j]}\dot{B}_{j[0i]}
-\frac{\dot{B}_{i[0i]}\dot{B}_{j[0j]}}{2(D-1)}-2\dot{h}_{ij}[\partial_{i}h_{0j}-\delta_{ij}\partial_{k}h_{0k}] \nn\\
&& - \dot{B}_{i[0j]}\Big[\partial_{j}B_{0[0i]}-\frac{\delta_{ij}\partial_{k}B_{0[0k]}}{(D-1)}\Big]
- \dot{B}_{i[0j]}\Big[\partial_{k}B_{j[ki]} -\frac{\delta_{ij}\partial_{k}B_{l[kl]}}{(D-1)}\Big]\nn\\
&& +m\dot{B}_{0[0i]}e_{i0} - m\dot{B}_{i[0j]}e_{ji} - \mathcal{V} \;,
\eea
where $h_{\mu\nu}=e_{(\mu\nu)}$ and the potential $\mathcal{V}$ has no time derivative. From the canonical momenta
$\pi^{\mu\nu}= \delta\mathcal{L}_{BT}/\delta\dot{e}_{\mu\nu}$ and $\Pi^{\mu[\alpha\beta]}=\delta\mathcal{L}_{BT}/\delta\dot{B}_{\mu[\alpha\beta]}$
we have the following primary constraints

\bea
\varphi^{00} &=& \pi^{00} \approx 0\;; \label{PC-1}\\
\varphi^{0i} &=& \pi^{0i} \approx 0\;; \label{PC-2}\\
\varphi^{i0} &=& \pi^{i0} \approx 0\;; \label{PC-3}\\
\varphi^{[ij]} &=& \pi^{[ij]} \approx 0\;; \label{PC-4}\\
\Phi^{0[ij]} &=& \Pi^{0[ij]} \approx 0\;; \label{PC-5}\\
\Phi^{i[jk]} &=& \Pi^{i[jk]} \approx 0\;; \label{PC-6}\\
\Phi^{0[0i]} &=& \Pi^{0[0i]} - m\,e_{i0} \approx 0\;; \label{PC-7}\\
\Phi^{i[0i]} &=& \Pi^{i[0i]}+m\,h_{ii} \approx 0\;. \label{PC-8}
\eea

The primary Hamiltonian is defined by

\bea
H_{P} = \int{d}^{D-1}x\Big[\mathcal{H}+\lambda_{\mu\nu}\varphi^{\mu\nu}+\Lambda_{\mu[\alpha\beta]}\Phi^{\mu[\alpha\beta]}\Big]\;,
\label{primary-s2}\eea
where $\lambda_{\mu\nu}$ and $\Lambda_{\mu[\alpha\beta]}$ are Lagrange multipliers such that $\lambda_{ij} = \lambda_{[ij]}$ and $\Lambda_{i[0i]}=\delta_{ij}\Lambda_{i[0j]}$, and the canonical Hamiltonian is:

\bea
\mathcal{H} &=&
\frac{1}{2}\pi^{(ij)}\pi^{(ij)}-\frac{\pi^{(ii)}\pi^{(jj)}}{2(D-2)}+\frac{1}{2}\Pi^{i[0j]}\Pi^{j[0i]}+2\pi^{(ij)}\partial_{i}h_{0j}+\Pi^{i[0j]}(\partial_{i}B_{0[0j]}+\partial_{k}B_{i[kj]}+m\,e_{ij})\nn\\
&&+\frac{1}{2}\partial_{i}h_{jk}\partial_{i}h_{jk}+\partial_{i}h_{00}(\partial_{i}h_{jj}-\partial_{j}h_{ij})-\frac{1}{2}\partial_{i}h_{jj}(\partial_{i}h_{kk}-2\partial_{k}h_{ik})-\partial_{i}h_{ij}\partial_{k}h_{kj}\nn\\
&&+\frac{m^{2}}{2}e_{ij}e_{ji}-\partial_{i}B_{0[ij]}\partial_{k}B_{j[0k]}-mB_{0[0i]}(\partial_{i}h_{00}+\partial_{j}e_{ij})-mB_{0[ij]}\partial_{i}e_{j0}+mB_{i[0j]}\partial_{j}e_{0i}\;.\nn\\
\label{density-s2}\eea

Applying the consistency condition to the primary constraints $\varphi^{i0}\approx0$ and $\Phi^{0[0i]}\approx0$ we obtain the Lagrange multipliers $\Lambda_{0[0i]}$ and $\lambda_{i0}$ respectively

\bea
\dot\varphi^{i0} &\approx& 0 \qquad \rightarrow \qquad \Lambda_{0[0i]} \approx -\,\frac{\partial_{j}\pi^{(ij)}}{m}+\partial_{j}B_{0[ji]}\;,\\
\dot\Phi^{0[0i]} &\approx& 0 \qquad \rightarrow \qquad \lambda_{i0} \approx \frac{\partial_{j}\Pi^{j[0i]}}{m}+\partial_{i}h_{00}+m\partial_{j}e_{ij}\;.
\eea

Back with these results in (\ref{primary-s2}) and applying the consistency condition to the other primary constraints we obtain the secondary constraints

\bea
\dot\varphi^{00} \approx0 \quad &\rightarrow& \quad
\chi = \partial_{i}\pi^{i0}+\nabla^{2}h_{ii}-\partial_{i}\partial_{j}h_{ij}-m\partial_{i}B_{0[0i]} \approx 0 \;; \label{SC-1}\\
\dot\varphi^{0i} \approx0 \quad &\rightarrow& \quad
\chi^{i} = \partial_{j}(\pi^{(ij)}-\pi^{[ij]}+mB_{i[0j]}) \approx 0 \;; \label{SC-2}\\
\dot\varphi^{[ij]} \approx0 \quad &\rightarrow& \quad
\chi^{[ij]} = m\Pi^{i[0j]}+\partial_{i}(\pi^{j0}-mB_{0[0j]})-m^{2}e_{ij} - (i\leftrightarrow{j})\approx 0 \;; \label{SC-3}\\
\dot\Phi^{i[0i]}\approx0 \quad &\rightarrow& \quad
\Theta = \partial_{i}\Pi^{0[0i]}-\frac{m\,\pi^{kk}}{(D-2)}\approx0 \;; \label{SC-4}\\
\dot\Phi^{0[ij]}\approx0 \quad &\rightarrow& \quad
\Theta^{[ij]} = \partial_{i}(\Pi^{0[0j]}-\partial_{k}B_{j[0k]}) - (i\leftrightarrow{j})\approx0 \;; \label{SC-5}\\
\dot\Phi^{i[jk]}\approx0 \quad &\rightarrow& \quad
\Theta^{i[jk]} = \partial_{j}\Pi^{i[0k]} - \partial_{k}\Pi^{i[0j]} \approx 0 \;. \label{SC-6}
\eea

Following the Dirac-Bergmann procedure, it turns out that  the consistency conditions for the secondary constraints:
$\dot{\chi}\approx0$, $\dot{\chi}^{i}\approx0$, $\dot{\Theta}^{[ij]}\approx0$ and $\dot{\Theta}^{i[jk]}\approx0$ are identically satisfied while from $\dot{\chi}^{[ij]}\approx0$ and $\dot{\Theta}\approx0$ we obtain the Lagrange multipliers $\lambda_{[ij]}$ and $\Lambda_{i[0i]}$ respectively. Therefore, there are no tertiary constraints in the model and the algorithm ends.

The classification of the constraints in first and second-class is obtained by calculating their Poisson brackets.
The constraints ($\varphi^{00}\,,\,\varphi^{0i}\,,\,\Phi^{0[ij]}\,,\,\Phi^{i[jk]}\,,\,\chi\,,\,\chi^{i}\,,\,\Theta^{[ij]}\,,\,\Theta^{i[jk]}$)
are first class since the Poisson bracket with all constraints are weakly zero. They correspond respectively to
$(1,\,D-1,\,(D-1)(D-2)/2,\,(D-1)^{2}(D-2)/2,\,1,\,D-1,\,D-2,\,(D-1)(D-2))$ constraints\footnote{The constraints $\Theta^{[ij]}=0$ have $D-2$ independent components as explained in footnote 3. The tensor $\Theta^{i[jk]}$ has in general $(D-1)^{2}(D-2)/2$ components however, the constraints $\Theta^{i[jk]}=0$ lead to $\Pi^{i[0j]}=\partial^{j}\phi^{i}$ which means that we have $(D-1)^{2}-(D-1)=(D-1)(D-2)$ independent constraints.}
such that the model has in total $(D^{3}-D^{2}+2D)/2$ first class constraints (FCC).

On the other hand, the constraints ($\pi^{i0}\,,\,\pi^{[ij]}\,,\,\Phi^{0[0i]}\,,\,\Phi^{i[0i]}\,,\,\chi^{[ij]}\,,\,\Theta$)
are second class since

\bea
\{\varphi^{i0} (y)\,,\,\Phi^{0[0j]}(x)\} &=& m\,\delta_{ij}\,\delta^{D-1}(x-y) \;;\\
\{\varphi^{[ij]} (y)\,,\,\chi^{[kl]}\,(x)\} &=& m^{2}(\delta_{ik}\delta_{jl}-\delta_{il}\delta_{jk})\,\delta^{D-1}(x-y)\;; \\
\{\Phi^{i[0i]} (y)\;,\,\Theta\;(x)\} &=& -\,\frac{m^{2}(D-1)}{(D-2)}\,\delta^{D-1}(x-y) \;. \eea
Thus, we have $(D^{2}-D+2)$ second class constraints (SCC).

The action (\ref{BT}) describes a massive spin-2 particle in terms of fields $e_{\mu\nu}$ and $B^{\mu[\alpha\beta]}$. In $D$ dimensions the field $e_{\mu\nu}$ has $D^{2}$ components while $B^{\mu[\alpha\beta]}$ has $D^{2}(D-1)/2$ components. In the phase space this corresponds to $N=D^{2}(D+1)$ components. The number of physical degrees of freedom ($dof$) in the phase space is given by

\bea
dof &=& N - 2\times{FCC} - SCC \nn\\
&=& (D+1)(D-2) \;.
\eea
Thus, in $D=3+1$ we have five Lagrangian $dof$ corresponding to helicities $\pm2$, $\pm1$ and $0$ as expected. In $D=2+1$ we have two $dof$ corresponding to helicities $\pm2$.

\subsection{Hamiltonian positivity}

We begin the positivity analysis by replacing the constraints as strong equalities in the primary Hamiltonian (\ref{primary-s2}). The partially reduced Hamiltonian $H^{(pr)}$ is given by

\bea
H^{(pr)} &=& H_{1} + H_{2}\;,
\eea
where

\bea
H_{1} &=& \int{d}^{D-1}x\Bigg\{\frac{1}{2}\pi^{(ij)}\pi^{(ij)}-\frac{1}{2}\frac{\pi^{jj}\pi^{kk}}{(D-2)}+\pi^{(ij)}\partial_{i}e_{j0}\Bigg\}\;,\label{H1}\\
H_{2} &=& \int{d}^{D-1}x\Bigg\{\frac{1}{2}\Pi^{i[0j]}\Pi^{j[0i]}+\Pi^{i[0j]}\partial_{i}B_{0[0j]}+m\Pi^{i[0j]}h_{ij}+m\,h_{ij}\partial_{i}B_{0[0j]}+\frac{m^{2}}{2}e_{[ij]}e_{[ij]}\nn\\
&&\qquad\quad\quad+\frac{m^{2}}{2}h_{ij}h_{ij}+\frac{1}{2}\partial_{i}h_{jk}\partial_{i}h_{jk}-\frac{1}{2}\partial_{i}h_{jj}\partial_{i}h_{kk}+\partial_{i}h_{jj}\partial_{k}h_{ik}-\partial_{i}h_{ij}\partial_{k}h_{kj}\Bigg\}\;.
\label{H2}\eea

From the constraint $\Theta^{[ij]}=0$ we have

\bea
&&\partial_{i}\Pi^{0[0j]}-\partial_{j}\Pi^{0[0i]} = \partial_{i}\partial_{k}B_{j[0k]}-\partial_{j}\partial_{k}B_{i[0k]}\;. \label{EQ-11}
\eea
Applying $\frac{\pi^{(jk)}\omega_{ki}}{m}$ in (\ref{EQ-11}) and using the constraints $\Phi^{0[0i]}=0$, $\varphi^{[ij]}=0$, $\chi^{i}=0$ and $\Theta=0$ we obtain (up to total integration) \footnote{The spin-0 projection operator $\omega_{ij}$ acting on Euclidean vectors as well as other projection operators are defined in the appendix.}

\bea
\pi^{(ik)}\partial_{i}e_{j0} = \frac{\pi^{(ik)}\omega_{ik}\pi^{jj}}{(D-2)}+\frac{\pi^{(ik)}\nabla^{2}\omega_{ik}\omega_{jl}\pi^{(jl)}}{m^{2}}
-\frac{\pi^{(ik)}\nabla^{2}\omega_{kj}\pi^{(ij)}}{m^{2}}\;.
\label{EQ-A}\eea
Back in (\ref{H1}) we rewrite $H_{1}$ in terms of $\pi^{(ij)}$

\bea
H_{1} &=& \int{d}^{D-1}x\Bigg\{\frac{1}{2}\pi^{(ij)}\pi^{(ij)}-\frac{\pi^{jj}\pi^{kk}}{2(D-2)}+\frac{\pi^{(ik)}\omega_{ik}\pi^{jj}}{(D-2)}+\frac{\pi^{(ik)}\nabla^{2}\omega_{ik}\omega_{jl}\pi^{(jl)}}{m^{2}}\nn\\
&&\qquad\qquad\quad-\,\frac{\pi^{(ik)}\nabla^{2}\omega_{kj}\pi^{(ij)}}{m^{2}}\Bigg\}\;.
\eea

We can rewrite the bilinear contractions of rank-2 tensors and their derivatives in terms of
spin projection and transition operators (see the Appendix). So

\bea
H_{1} &=&
\frac{1}{2}\int{d}^{D-1}x\,\pi^{(ij)}\Bigg[P^{(2)}_{SS}+\frac{(m^{2}-\nabla^{2})}{m^{2}}P^{(1)}_{SS}+\frac{(D-1)}{(D-2)}P^{(0)}_{WW}\Bigg]_{ij,kl}\pi^{(kl)}\;,
\label{Hr-posit}\eea
which is clearly positively defined for $D>2$. The positivity of $H_{1}$ is clear since any quadratic form with a sandwiched projection operator is non negative in Euclidean space, $\Pi^A P_{AB} \Pi^B = (P_{CA} \Pi^A)((P_{CB} \Pi^B)\ge 0$. Each helicity contribution is separately positive definite.

Let us now check the positivity of $H_{2}$.  From the the constraints $\Theta^{i[ik]}=0$ and $\Phi^{i[0i]}=0$ we find the following relation

\bea
\partial_{j}\Pi^{j[0k]} = -\,m\partial_{k}h_{jj}\;. \label{EQ-21}
\eea

If we apply $\frac{\Pi^{k[0i]}\partial_{j}}{\nabla^{2}}$ on $\Theta^{i[jk]}=0$ and use (\ref{EQ-21}) we obtain (after some total integrations)

\bea
\Pi^{k[0i]}\Pi^{i[0k]} = m^{2}h_{ii}h_{kk}\;. \label{EQ-B}
\eea
On the other hand, if we apply $B_{0[0k]}\omega_{ij}$ on $\Theta^{i[jk]}=0$ and use (\ref{EQ-21}) we find

\bea
B_{0[0k]}\partial_{i}\Pi^{i[0k]} = m\,h_{ii}\partial_{k}B_{0[0k]}\;. \label{EQ-23}
\eea

From the constraints $\varphi^{i0}=0$ and $\chi=0$ we have

\bea
m\partial_{i}B_{0[0i]} = \nabla^{2}h_{ii}-\partial_{i}\partial_{j}h_{ij}\;. \label{EQ-24}
\eea
Substituting (\ref{EQ-24}) in (\ref{EQ-23}) we obtain (up to total integration)

\bea
\Pi^{i[0k]}\partial_{i}B_{0[0k]} = -\,h_{ii}\nabla^{2}h_{kk}+h_{ii}\partial_{j}\partial_{k}h_{jk}\;. \label{EQ-C}
\eea
If we apply $\frac{m\,h_{ik}\partial_{j}}{\nabla^{2}}$ in $\Theta^{i[jk]}=0$ we have

\bea
m\,h_{ik}\Pi^{i[0k]} = m\,h_{ik}\omega_{kj}\Pi^{i[0j]}\;. \label{EQ-D}
\eea

\no Now consider the constraints $\varphi^{i0}=0$ and $\chi^{[ij]}=0$. So

\bea
m\partial_{i}B_{0[0j]}-m\partial_{j}B_{0[0i]} = m\Pi^{i[0j]}-m\Pi^{j[0i]}-2m^{2}e_{[ij]} \;. \label{EQ-25}
\eea
Applying $h_{jk}\omega_{ki}$ on (\ref{EQ-25}) and using (\ref{EQ-21}) and (\ref{EQ-24}) we obtain

\bea
m\,h_{jk}\partial_{k}B_{0[0j]} &=&
-\,m\,h_{jk}\omega_{ki}\Pi^{j[0i]}-m^{2}h_{ii}\omega_{jk}h_{jk}+h_{ii}\nabla^{2}\omega_{jk}h_{jk} \nn\\
&&-h_{ij}\nabla^{2}\omega_{ij}\omega_{kl}h_{kl}+2m^{2}e_{[ij]}\omega_{jk}h_{ik} \;. \label{EQ-E}
\eea

Finally, substituting the relations (\ref{EQ-B}), (\ref{EQ-C}), (\ref{EQ-D}) and (\ref{EQ-E}) in $H_{2}$ we obtain

\bea
H_{2} &=& \int{d}^{D-1}x\Bigg\{-\frac{1}{2}h_{ij}\nabla^{2}h_{ij}-\frac{1}{2}h_{ii}\nabla^{2}h_{jj}+h_{ii}\nabla^{2}\omega_{jk}h_{jk}+h_{ij}\nabla^{2}\omega_{jk}h_{ik}-m^{2}h_{ii}\omega_{jk}h_{jk}\nn\\
&&\qquad\qquad-\,h_{ij}\nabla^{2}\omega_{ij}\omega_{kl}h_{kl}+\frac{m^{2}}{2}h_{ij}h_{ij}+\frac{m^{2}}{2}h_{ii}h_{jj}+\frac{m^{2}}{2}e_{[ij]}e_{[ij]}+2m^{2}e_{[ij]}\omega_{jk}h_{ik}\Bigg\}\;,
\nn\eea
or

\bea
H_{2} &=& \int{d}^{D-1}x\Bigg\{-\frac{1}{2}h_{ij}\nabla^{2}h_{ij}-\frac{1}{2}h_{ii}\nabla^{2}h_{jj}+(\nabla^{2}-m^{2})h_{ii}\omega_{jk}h_{jk}+(\nabla^{2}-m^{2})h_{ij}\omega_{jk}h_{ik}\nn\\
&&\qquad\qquad-(\nabla^{2}-m^{2})h_{ij}\omega_{ij}\omega_{kl}h_{kl}+\frac{m^{2}}{2}h_{ij}h_{ij}+\frac{m^{2}}{2}h_{ii}h_{jj}+\frac{m^{2}}{2}E_{[ij]}E_{[ij]}\Bigg\}\;,\nn\\
\eea
where $E_{[ij]} = e_{[ij]}-\omega_{ik}h_{jk}+\omega_{jk}h_{ik}$. In terms of spin projection and transition operators (see Appendix) we can write

\bea
H_{2} &=& \frac{1}{2}\int{d}^{D-1}x\Bigg\{h_{ij}\Big[(m^{2}-\nabla^{2})P^{(2)}_{SS}+(D-1)(m^{2}-\nabla^{2})P^{(0)}_{SS}\Big]_{ij,kl}h_{kl}\nn\\
&&\qquad\qquad\quad + \frac{m^{2}}{2}E_{[ij]}\Big[P^{(1)}_{AA}+P^{(0)}_{AA}\Big]_{ij,kl}E_{[kl]}\Bigg\}\;.
\eea
This results is clearly positively defined. Therefore, the positivity of each spin mode contribution is demonstrated in terms of projection operators.

\section{Conclusions}

Here we have applied the plain Dirac-Bergmann algorithm in a new massive spin-2 model proposed in \cite{dual}. In the section 2 we have reviewed the case of the massive spin-1 abelian BF model
(in $D$ dimensions) as an introduction. We suggest a parent action that guarantees the duality between the BF and the Proca models.
The correlation functions in the Stueckelberg-like Proca model are mapped into correlation functions of the BF model.
The constraints analysis of the BF model shows that all constraints are first class.
We have found $(D-1)$ degrees of freedom which in $D=3+1$ corresponds to helicities $\pm1,\,0$.
The positivity of the Hamiltonian is confirmed using the spin-0 and spin-1 projection operators.
This procedure is different from that of \cite{Lahiri} where a gauge fixing was adopted as a strategy for reduction of phase space.

In the section 3 we have moved to the spin-2 case.
We have proposed a parent action that allows us an alternative proof of duality between the Fierz-Pauli (FP) and the new spin-2 model. The equivalence between correlation functions is obtained and the dual map between the two models is established in terms of gauge invariant quantities (up to contact terms). We have noticed that in the special case of $D=2+1$ the new massive model is a kind of ``zwei-dreibein'' model \cite{zwei}. Applying the Dirac procedure in the analysis of a dual massive spin-2 model we have found all the Hamiltonian constraints of the model.
Unlike the spin-1 BF model, the new spin-2 model has first- and second-class constraints. This is due to the more complicated tensor structure of the model.
We have found five degrees of freedom in $D=3+1$, which is compatible with a massive spin-2 particle.

Fixing the constraint strongly to zero in the primary Hamiltonian we obtain the reduced Hamiltonian. Using the spin projection operators is possible to write the reduced Hamiltonian in terms of a sum of squares allowing to check the Hamiltonian positivity.
Besides, it is possible to identify the contribution of each helicity separately.
Therefore, we conclude that the canonical structure of the phase space is faithful to massive spin-2 particles so that the dual spin-2 model can be used as an alternative starting point for building up diffeomorphism invariant massive gravities. We must now investigate the definition of possible cubic vertices consistent with the corresponding non linear extension of the local reparametrization symmetry.

\section{Acknowledgements}

A.L.R dos S. is supported by CNPq (grant 160784/2019-0); D.D. is partially supported by CNPq  (grant 306380/2017-0) while W.d.P. acknowledges the partial support of CNPQ under Grants No. 438562/2018-6 and No. 313236/2018-6, and the partial support of CAPES under Grant No. 88881.309870/2018-01.

\section{Appendix}

In $D$ dimensions the spin-0 and spin-1 projection operators acting on vector fields are defined in the spacial sector by

\be \omega_{ij}=\frac{\partial_{i}\partial_{j}}{\nabla^{2}}\qquad \mbox{and}\qquad \theta_{ij}=\delta_{ij}-\omega_{ij}\;,\ee
respectively. Using this projectors one can define the projection and transition operators. The symmetric operators are given by

\bea \Big(P^{(2)}_{SS}\Big)^{ij,kl}&=&\frac{1}{2}(\theta^{ik}\theta^{jl}+\theta^{il}\theta^{jk})-\frac{\theta^{ij}\theta^{kl}}{(D-2)}\;,\label{Pss2}\\
\Big(P^{(1)}_{SS}\Big)^{ij,kl}&=&\frac{1}{2}(\theta^{ik}\omega^{jl}+\theta^{il}\omega^{jk}+\theta^{jk}\omega^{il}+\theta^{jl}\omega^{ik})\;,\label{Pss1}\\
\Big(P^{(0)}_{SS}\Big)^{ij,kl}&=&\frac{\theta^{ij}\theta^{kl}}{D-2}\quad,\quad
\Big(P^{(0)}_{WW}\Big)^{ij,kl}=\omega^{ij}\omega^{kl}\;,\nn\\
\Big(P^{(0)}_{SW}\Big)^{ij,kl}&=&\frac{\theta^{ij}\omega^{kl}}{\sqrt{(D-2)}}\quad,\quad
\Big(P^{(0)}_{WS}\Big)^{ij,kl}=\frac{\omega^{ij}\theta^{kl}}{\sqrt{D-2}}\;,\nn\\
\eea
and satisfy the symmetric closure relation

\bea
\Big[P^{(2)}_{SS}+P^{(1)}_{SS}+P^{(0)}_{SS}+P^{(0)}_{WW}\Big]^{ij,kl}=\frac{1}{2}(\delta^{ik}\delta^{jl}+\delta^{il}\delta^{jk})\;.\label{symmClosure}
\eea

The antisymmetric and mixed operators are given by

\bea
\Big(P^{(1)}_{AA}\Big)^{ij,kl}&=&\frac{1}{2}(\theta^{ik}\omega^{jl}-\theta^{il}\omega^{jk}-\theta^{jk}\omega^{il}+\theta^{jl}\omega^{ik})\;,\label{Paa1}\\
\Big(P^{(0)}_{AA}\Big)^{ij,kl}&=&\frac{1}{2}(\theta^{ik}\theta^{jl}-\theta^{il}\theta^{jk})\label{P0}\;,\nn\\
\Big(P^{(1)}_{SA}\Big)^{ij,kl}&=&\frac{1}{2}(\theta^{ik}\omega^{jl}-\theta^{il}\omega^{jk}+\theta^{jk}\omega^{il}-\theta^{jl}\omega^{ik})\;,\label{Paa1}\\
\Big(P^{(1)}_{AS}\Big)^{ij,kl}&=&\frac{1}{2}(\theta^{ik}\omega^{jl}+\theta^{il}\omega^{jk}-\theta^{jk}\omega^{il}-\theta^{jl}\omega^{ik})\;,\label{Paa1}\\
\eea
and satisfy the antisymmetric closure relation

\bea
\Big[P^{(1)}_{AA}+P^{(0)}_{AA}\Big]^{ij,kl}=\frac{1}{2}(\delta^{ik}\delta^{jl}-\delta^{il}\delta^{jk})\;.\label{AntisymmClosure}
\eea

From (\ref{symmClosure}) and (\ref{AntisymmClosure}) we have the general closure relation

\bea
\Big[P^{(2)}_{SS}+P^{(1)}_{SS}+P^{(1)}_{AA}+P^{(0)}_{SS}+P^{(0)}_{WW}+P^{(0)}_{AA}\Big]^{ij,kl}=\delta^{ik}\delta^{jl}\;.
\eea

The operators satisfy the algebra
\bea
P^{(r)}_{IJ}P^{(s)}_{KL} = \delta^{rs}\delta_{JK}P^{(r)}_{IL}\;.
\eea

\end{document}